\documentclass[aps,PRB,twocolumn,groupedaddress,floats]{revtex4}
\usepackage{amssymb}
\usepackage{graphicx}

\begin{document}

\title{The orbital characters of bands in iron-based superconductor BaFe$_{1.85}$Co$_{0.15}$As$_2$}

\author{Y. Zhang, F. Chen, C. He, B. Zhou, B. P. Xie,}
\affiliation{State Key Laboratory of Surface Physics,  Key Laboratory of Micro
and Nano Photonic Structures (MOE), and Department of Physics, Fudan
University, Shanghai 200433, People's Republic of China}

\author{C. Fang, W. F. Tsai}

\affiliation{Department of Physics, Purdue University, West Lafayette, Indiana
47907, USA}

\author{X. H. Chen}

\affiliation{Department of Physics, University of Science and Technology of
China, Hefei, Anhui 230027, People's Republic of China}

\author{H. Hayashi, J. Jiang, H. Iwasawa, K. Shimada, H. Namatame, M.
Taniguchi}

\affiliation{Hiroshima Synchrotron Radiation Center and Graduate School of
Science, Hiroshima University, Hiroshima 739-8526, Japan.}

\author{J. P. Hu}\email{hu4@purdue.edu}
\affiliation{Department of Physics, Purdue University, West Lafayette, Indiana
47907, USA}

\author{D. L. Feng}\email{dlfeng@fudan.edu.cn}
\affiliation{State Key Laboratory of Surface Physics,  Key Laboratory of Micro
and Nano Photonic Structures (MOE), and Department of Physics, Fudan
University, Shanghai 200433, People's Republic of China}

\begin{abstract}

The unconventional superconductivity in the newly discovered iron-based
superconductors is intimately related to its multi-band/multi-orbital nature.
Here we report  the comprehensive orbital characters of the low-energy
three-dimensional electronic structure in BaFe$_{1.85}$Co$_{0.15}$As$_2$ by
studying the polarization and photon energy dependence of angle-resolved
photoemission data. While the distributions of the $d_{xz}$, $d_{yz}$, and
$d_{3z^2-r^2}$ orbitals agree with the prediction of density functional theory,
those of  the $d_{xy}$ and $d_{x^2-y^2}$ orbitals show remarkable disagreement
with theory. Our results point out the inadequacy of the existing band
structure calculations, and more importantly, provide a foundation for
constructing the correct microscopic model of iron pnictides.
\end{abstract}


\maketitle

\section{Introduction}

Unlike the cuprates, the low-energy electronic structure of the iron-based
superconductors is dominated by  multiple bands and orbitals
\cite{LDA1,Kuroki,Graser1}. Theoretically, it has been proposed that the Fermi
surface sheets with multiple orbitals could result in a strong anisotropy and
amplitude variation of the superconducting gaps \cite{DHLee,Ronny}.
Experimentally, a recent study on Ba$_{0.6}$K$_{0.4}$Fe$_2$As$_2$ shows that
the superconducting gap sizes are different at the same Fermi momentum for two
bands with different orbital characters\cite{YZhangBK}. To construct correct
models for the iron-based superconductors, and to understand the unconventional
superconductivity,  it is thus critical to experimentally identify the orbital
characters of the low-energy electronic structure.

There have been inconsistencies over the orbital identities of the bands
amongst theories \cite{Kuroki,Graser1}, and a few experiments
\cite{Hasan,ZXShen,Shin,Malaeb,Borisenko,Fink}. Moreover, various physical
properties of iron-based superconductors are featured with three dimensional
(3D) characters \cite{HQYuan}. For example, the band structure in the so-called
``122" series of iron-pnictides is rather 3D \cite{Malaeb, BaCo3D}, and the gap
dependency on the out-of-plane momentum ($k_z$) has recently been reported
\cite{YZhangBK}. It has been pointed out that the three dimensionality of the
electronic structure might be essential in inducing both the spin density wave
(SDW) and superconductivity\cite{Graser2}. However, the orbital character
distribution along the $k_z$ direction in 3D momentum space has not been
exposed so far.

Here we report a systematic  angle-resolved photoemission spectroscopy (ARPES)
study on the orbital character of the electronic structure in an electron-doped
``122" compound, BaFe$_{1.85}$Co$_{0.15}$As$_2$. We have observed strong
polarization dependency for all the bands near the Fermi energy ($E_F$), and
have obtained a comprehensive picture of the orbital characters of these bands.
We confirm previous theoretical findings that the bands with $d_{xz}$, $d_{yz}$
orbital form two hole pockets around the zone center and one electron pocket
around the zone corner, and the $d_{3z^2-r^2}$ (or $d_{z^2}$ for simplicity)
orbital  is mostly included in the two bands well below $E_F$. More
importantly, we find that certain bands predicted to be of the $d_{xy}$ orbital
in theory are actually mixed with the $d_{x^2-y^2}$ orbital around the zone
center, and are dominated by the $d_{x^2-y^2}$ orbital around the zone corner.
Furthermore, we have identified the orbital characters of the band structure in
3D momentum space. Our results provide explicit knowledge for constructing the
theoretical model of iron-based superconductors.

\section{Polarization-dependent ARPES}

\begin{figure}
\centerline{\includegraphics[width=8.7cm]{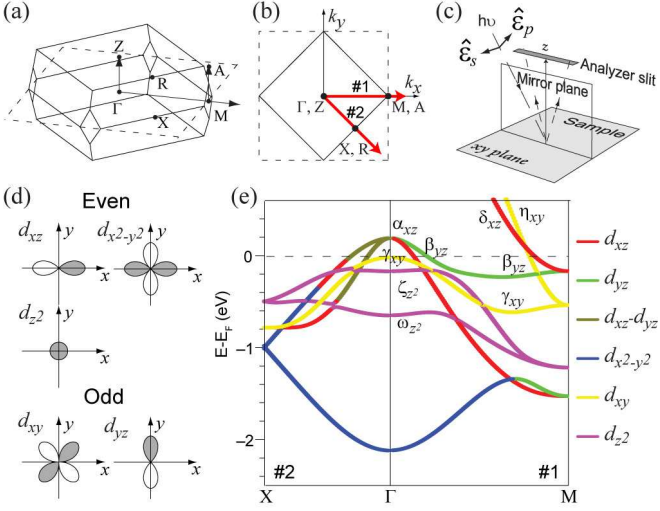}} \caption{(Color
online) The experimental setup and definitions. (a) The Brillouin
zone of BaFe$_{1.85}$Co$_{0.15}$As$_2$. (b) Two-dimensional plot of
the simplified Brillouin zone (solid line) and the unfolded
Brillouin zone (dashed line). (c)  Experimental setup for
polarization-dependent ARPES. For the $p$ (or $s$) experimental
geometry, the electric field direction of the incident photons
$\bf{\hat{\varepsilon}_{p}}$ (or $\bf{\hat{\varepsilon}_{s}}$)  is
parallel (or perpendicular) to the mirror plane defined by the
analyzer slit and the sample surface normal. (d) Illustration of the
spatial symmetry of the $3d$ orbitals with respect to the mirror
plane formed by surface normal and cut \#1 in panel b, \textit{i.e.}
the $xz$ plane. (e) A typical orbital assignment of bands of iron
pnictide as calculated in Ref.~[3] .} \label{setup}
\end{figure}

BaFe$_{1.85}$Co$_{0.15}$As$_2$ single crystals were synthesized by a self-flux
method \cite{XHChen} with a superconducting transition temperature ($T_c$) of
25~K, and no SDW or structural transition. As this system is optimally  doped
with electrons, the low energy band structure is well occupied, and free from
the complications of the electronic reconstruction in the SDW state
\cite{LXYang, YZhang, Myi, Che}. Because the arsenic ions in the FeAs layer are
situated in two inequivalent positions, there are two iron ions per unit cell.
The Brillouin zone of BaFe$_{1.85}$Co$_{0.15}$As$_2$ is shown in
Figs.~\ref{setup}(a) and ~\ref{setup}(b). To compare with theory, we define the
$k_x$ and $k_y$ directions to be the Fe-Fe bond directions, as shown in the
unfolded Brillouin zone (dashed lines) for one iron ion per unit cell. Two high
symmetry directions, $\Gamma$(Z)-M(A) and $\Gamma$(Z)-X(R), are illustrated by
red arrows \#1 and \#2, respectively in Fig.~\ref{setup}(b).

The polarization-sensitivity of the orbitals in ARPES is a powerful tool to
identify the orbital characters of band structure \cite{ZXShenRev}. The matrix
element of the photoemission process can be described by
$${|M_{f,i}^{\bf{k}}|\propto{\rm{|}}\langle \phi _f^{\bf{k}}
|\bf{\hat{\varepsilon}}\cdot{\bf{r}}|\phi _i^{\bf{k}} \rangle |^2}$$, where
$\bf{\hat{\varepsilon}}$ is the unit vector of the electric field of the light
\cite{ZXShenRev}. For high kinetic-energy photoelectrons, the final-state
wavefunction ${\phi _f^{\bf{k}}}$ can be approximated by a plane-wave state
${e^{i{\bf{k}\cdot\bf{r}}}}$ with $\bf{k}$ in the mirror plane as plotted in
Fig.~\ref{setup}(c). Consequently, it is always even with respect to the mirror
plane. For the $p$ (or $s$) experimental geometry in Fig.~\ref{setup}(c),
because $\bf{\hat{\varepsilon}}$ is parallel (or perpendicular) to the mirror
plane, $\bf{\hat{\varepsilon}}\cdot{\bf{r}}$ is even (or odd). Therefore, to
have finite matrix element,  \textit{i.e.} to be observed, the initial state
$\phi _i^{\bf{k}}$ has to be even (or odd) in the $p$ (or $s$) geometry.

Considering the spatial symmetry of the $3d$ orbitals [Fig.~\ref{setup}(d)],
when the analyzer slit is along the high symmetry direction of the sample, the
photoemission signal of certain orbitals would appear or disappear by
specifying the polarization directions as summarized in Table~\ref{t1}. For
example, with respect to the mirror plane formed by  direction \#1 and sample
surface normal (or the $xz$ plane), the even orbitals ($d_{xz}$, $d_{z^2}$, and
$d_{x^2-y^2}$) and the odd orbitals ($d_{xy}$ and $d_{yz}$) could be only
observed in the $p$ and $s$ geometry respectively. Note that, $d_{xz}$ and
$d_{yz}$ are not symmetric with respect to the mirror plane defined by
direction \#2 and surface normal, thus could be observed in both the $p$ and
$s$ geometries.

\begin{table}
\caption{The possibility to detect $3d$ orbitals along two high symmetry
directions in the $p$ and $s$ geometry by polarization-dependent APRES}
\begin{tabular}{@{\vrule height 10.5pt depth 4pt  width0pt}ccccccc}
High-symmetry & Experimental &\multicolumn{5}{c}{$3d$ orbitals}\\
\cline{3-7} direction & geometry & $d_{xz}$ & $d_{x^2-y^2}$ &
$d_{z^2 }$&$d_{yz}$&$d_{xy}$\\ \hline
\#1 $\Gamma$(Z)-M(A)  & $p$ &$\surd$&$\surd$&$\surd$& & \\
& $s$ & & & &$\surd$&$\surd$ \\ \hline
\#2 $\Gamma$(Z)-X(R)  & $p$ &$\surd$& &$\surd$ &$\surd$ &$\surd$ \\
& $s$ &$\surd$&$\surd$& &$\surd$ & \\ \hline
\end{tabular}    \label{t1}
\end{table}

Figure~\ref{setup}(e) shows the band calculation of iron-based superconductors
in a two-dimensional (2D) band model reproduced from Ref.\cite{Graser1}, which
was confirmed by many other calculations, and widely adopted in various models.
The bands are labeled with corresponding orbital characters as subscripts.
Around the $\Gamma$ point, the $d_{z^2}$ orbital contributes to the two bands
$\zeta_{z^2}$ and $\omega_{z^2}$ well below $E_F$. There are three holelike
bands near $E_F$. $\alpha_{xz}$ and $\beta_{yz}$ cross $E_F$ forming two hole
pockets, while the band top of $\gamma_{xy}$ is just below $E_F$. The
$\beta_{yz}$ and $\gamma_{xy}$ disperse to the lower binding energies at M. Two
bands, $\delta_{xz}$ and $\eta_{xy}$, form two electron pockets, and they are
degenerate with $\beta_{yz}$ and $\gamma_{xy}$ at M respectively. The
$d_{x^2-y^2}$ orbital was found to be irrelevant to the low-energy electronic
structure.

ARPES measurements were performed at the Beamline 1 of Hiroshima synchrotron
radiation center (HSRC). By rotating the the entire photoemission spectroscope
around the incoming photon beam, one can switch between the $s$ and $p$
polarization geometries. All the data were taken with a Scienta electron
analyzer at 30~K, the overall energy resolution is 15~meV, and the angular
resolution is 0.3 $^{\circ}$. The sample was cleaved \textit{in situ}, and
measured under ultra-high-vacuum of $5\times10^{-11}$\textit{torr}.

\section{Results}

\subsection{Orbital characters of bands around the zone center}

\begin{figure}
\centerline{\includegraphics[width=8.7cm]{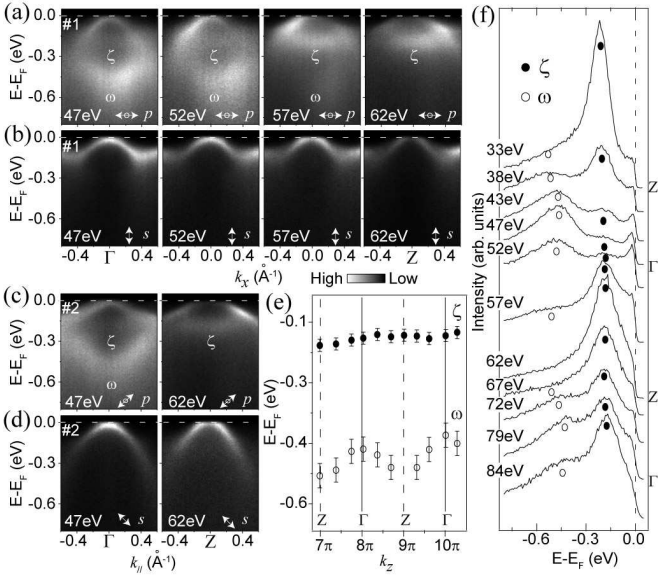}}
\caption{Photoemission data around the zone center. (a) and (b), The
photon energy dependence of the photoemission intensity
[$I(k,\omega)$] along $\Gamma$(Z)-M(A) (\#1) in the $p$ and $s$
geometries, respectively. (c) and (d), The photon energy dependence
of $I(k,\omega)$ along $\Gamma$(Z)-X(R) (\#1) in the $p$ and $s$
geometries, respectively. (e) The $k_z$ dispersion of the $\zeta$
and $\omega$ bands along $\Gamma$-Z. (f) The photon energy
dependence of EDCs sampled along $\Gamma$-Z. } \label{z2}
\end{figure}

The photoemission data taken near the zone center are shown in Fig.~\ref{z2}.
Two bands assigned as $\zeta$ and $\omega$ could be observed at about 170 and
500~meV below $E_F$. Both bands only emerge in the $p$ geometry, and could not
be observed in the $s$ geometry along both \#1 and \#2 directions
[Figs.~\ref{z2}(a)-\ref{z2}(d)]. Based on Table~\ref{t1}, these two bands are
made of $d_{z^2}$ orbital, assuming they are consisted of a single orbital as
suggested in Fig.~\ref{setup}(e). The energy distribution curves (EDCs) at
$k_{\parallel}~=~0~\AA^{-1}$ taken with different photon energies are stacked
in Fig.~\ref{z2}(f), so that we could track the $\zeta$ and $\omega$ bands
along $\Gamma$-Z. The peak position of $\omega$ shows a periodic variation if
we take the inner potential to be 15~eV [Fig.~\ref{z2}(e)]. Furthermore, the
peak intensities of $\omega$ and $\zeta$ show periodic anti-correlation with
the photon energy. While $\omega$ is at its strongest at $\Gamma$ (47 and
79~eV), $\zeta$ is mostly enhanced at Z (33 and 62~eV). The anti-correlation of
intensities of these two bands could be naturally explained by the different
wavefunction distributions along the $z$ direction for the bonding and
antibonding states formed by the $d_{z^2}$ orbitals \cite{bilayer}. Summing up
these facts, we attribute $\zeta$ and $\omega$ to $d_{z^2}$ orbital, which
agrees with the assignments in Fig.~\ref{setup}(e).

\begin{figure}
\centerline{\includegraphics[width=8.7cm]{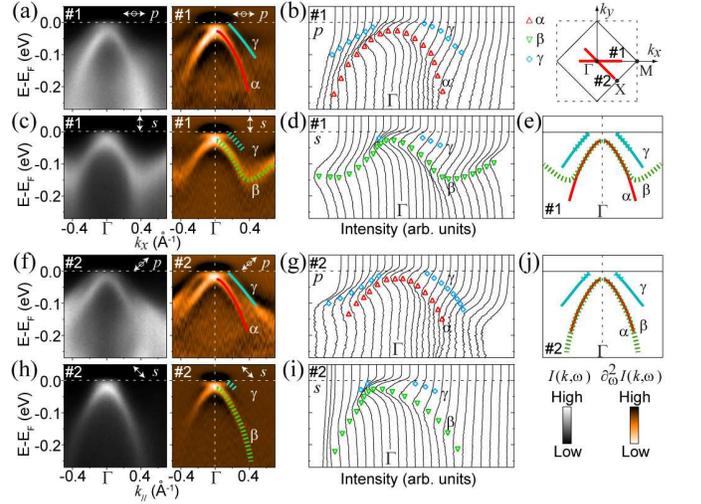}} \caption{(Color
online) The polarization dependent APRES data around $\Gamma$ taken
with 47~eV photons. (a) $I(k,\omega)$ map  and its second derivative
with respect to energy ($\partial^2 I(k,\omega)/\partial\omega^2$)
taken with the $p$ geometry along \#1 direction as marked with a red
thick solid line in the inset. (b) The EDCs for the data in panel a.
(c) and (d) are the same as  panels a and b, but taken with the $s$
geometry. (e) The sketch of the band structure in panels a and c.
(f)-(j) are the same as (a)-(e), but taken along cut \#2 in the
inset. The band dispersions are identified with the help of second
derivative image and EDCs, and illustrated by the solid and dashed
lines ploted on the second derivative image. The color scale of
$I(k,\omega)$ and $\partial^2 I(k,\omega)/\partial\omega^2$ are
shown in the inset. The solid  and dashed sketches of  the bands
represent the even and odd components of the bands respectively.
These color scales and notations are complied throughout the paper.}
\label{gamma}
\end{figure}

Near the Fermi energy, three bands ($\alpha$, $\beta$, and $\gamma$) could be
identified around $\Gamma$ in Fig.~\ref{gamma}. $\alpha$ only shows up in the
$p$ geometry, while $\beta$ only appears in the $s$ geometry, exhibiting
opposite spatial symmetries [Figs.~\ref{gamma}(a)-(e)]. The band tops of both
$\alpha$ and $\beta$ are below $E_F$ and degenerate at the $\Gamma$ point. Note
that the bands with $d_{xz}$ and $d_{yz}$ orbitals should be degenerate at the
$\Gamma$ point due to the four-fold symmetry of the sample. Therefore, we
attribute the $\alpha$ and $\beta$ bands to be of the $d_{xz}$ and $d_{yz}$
orbitals, respectively in this direction. In addition, along the $\Gamma$-M
direction, $\alpha$ disperses to the binding energy over 200~meV, and $\beta$
disperses to about 150~meV and then bends over to lower binding energies toward
the M point, which is consistent with the behavior of $\alpha_{xz}$ and
$\beta_{yz}$ predicted by theory in Fig.~\ref{setup}(e). The $d_{xz}$ and
$d_{yz}$ orbitals have no definite symmetry with respect to the $\Gamma$-X
direction. The strong polarization dependence of $\alpha$ and $\beta$ observed
along $\Gamma$-X direction [Figs.~\ref{gamma}(f)-\ref{gamma}(j)] are most
likely due to the hybridization of the $d_{xz}$ and $d_{yz}$ orbitals. For
example, $\frac{d_{xz}+d_{yz}}{\sqrt{2}}$ and $\frac{d_{xz}-d_{yz}}{\sqrt{2}}$
are of even and odd spatial symmetry with respect to the $\Gamma$-X direction,
respectively. Actually, because of the four-fold symmetry, the $d_{xz}$
component of $\alpha$ along the $k_x$ axis has to become $d_{yz}$ along the
$k_y$-axis. Therefore, the equal mixing of these two orbitals along the
$\Gamma$-X direction is expected. Similar arguments hold for the $\beta$ band.

The $\gamma$ band could be observed in both geometries, indicating that it is a
mixture of both odd and even orbitals. With respect to $\Gamma$-M, the odd
orbital candidates are $d_{xy}$ and $d_{yz}$. As the $d_{yz}$ orbital
contributes to $\beta$,  the odd orbital in $\gamma$ is likely $d_{xy}$, as the
$\gamma_{xy}$ band proposed  in Fig.~\ref{setup}(e). The even orbital
candidates include $d_{xz}$, $d_{z^2}$ and $d_{x^2-y^2}$. Since $\alpha$
($d_{xz}$), $\zeta$ ($d_{z^2}$), and $\omega$ ($d_{z^2}$) bands contains no odd
orbital, the odd orbitals  in $\gamma$ are less likely mixed with the $d_{xz}$
and $d_{z^2}$ orbital in this momentum region. We thus deduce that the most
possible even orbital in $\gamma$ is $d_{x^2-y^2}$. Because both $d_{xy}$ and
$d_{x^2-y^2}$ orbitals mainly spread in the FeAs plane,  the $\gamma$ band are
expected to be more two dimensional than the $\alpha$ and $\beta$ bands. Such
an assignment is consistent with the polarization dependence of data along
$\Gamma$-X as well [Figs.~\ref{gamma}(f)-\ref{gamma}(j)], just $d_{x^2-y^2}$ is
the odd orbital, and  $d_{xy}$ is the even one in this direction.

\begin{figure}
\centerline{\includegraphics[width=8.7cm]{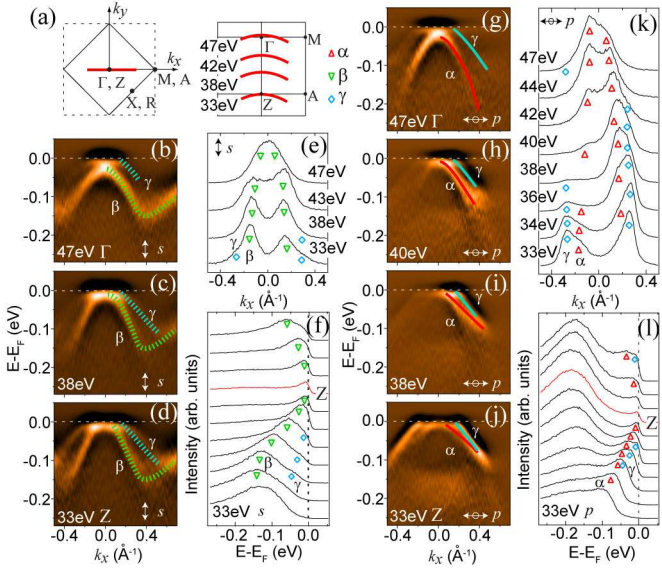}} \caption{(Color
online) Photon energy dependence of bands near the zone center. (a)
Illustration of the cuts sampled with different photon energies.
(b)-(d) $\partial^2 I(k,\omega)/\partial\omega^2$ taken in the $s$
geometry with 47, 38 and 33~eV photons respectively. (e) The photon
energy dependence of the MDCs at 30~meV below $E_F$ in the $s$
geometry. (f) The EDCs for the data in panel d. (g)-(j) $\partial^2
I(k,\omega)/\partial\omega^2$ taken in the $p$ geometry with 47, 40,
38 and 33~eV photons respectively. (k) The photon energy dependence
of the MDCs at 30~meV below $E_F$ in the $p$ geometry. (l) The EDCs
for the data in panel j.} \label{gphd}
\end{figure}

\begin{figure}
\centerline{\includegraphics[width=8.7cm]{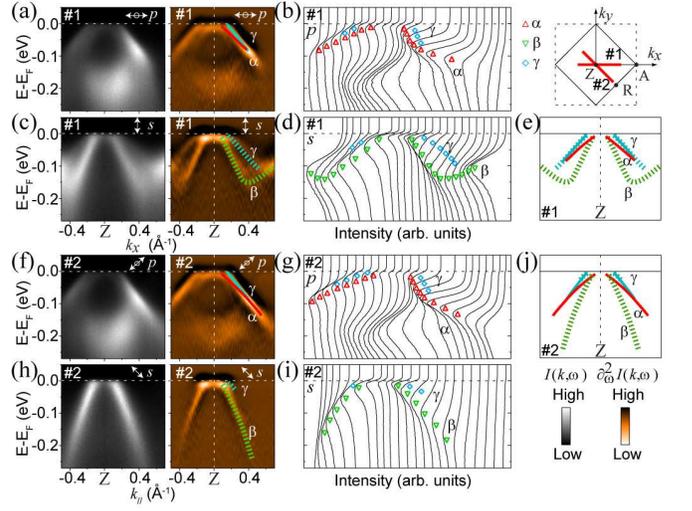}} \caption{(Color
online) The polarization dependent APRES data around Z taken with
33~eV photons. (a) $I(k,\omega)$ map  and $\partial^2
I(k,\omega)/\partial\omega^2$ taken with the $p$ geometry along \#1
direction as marked with a red thick solid line in the inset. (b)
The EDCs for the data in panel a. (c) and (d) are the same as in
panels a and b, but taken with the $s$ geometry. (e) The sketch of
the band structure in panels a and c. (f)-(j) are the same as in
(a)-(e), but taken along \#2 direction as marked with a red thick
solid line in the inset.} \label{z}
\end{figure}

To confirm such 2D nature of the $\gamma$ band, and explore the 3D
characters of the $\alpha$ and $\beta$ bands, photon energy
dependent data around the zone center are shown in Fig.~\ref{gphd}.
The $\beta$ band shifts away from the zone center when approaching
the Z point [Figs.~\ref{gphd}(b)-(d)], which is also illustrated by
the shift of the momentum distribution curve (MDC) peaks of $\beta$
in Fig.~\ref{gphd}(e). The $k_z$ dependence of the $\alpha$ band
shows similar behavior as $\beta$, with its band top below $E_F$ at
$\Gamma$, but slightly above $E_F$ at Z
[Figs.~\ref{gphd}(g)-\ref{gphd}(j)]. Moreover, the in-plane
dispersion of $\alpha$ becomes quite flat and eventually intersect
the $\gamma$ band around Z as shown in Figs.~\ref{gphd}(j) and
\ref{gphd}(l). The MDC peak positions of $\gamma$ show almost no
photon energy dependence in Fig.~\ref{gphd}(k), and no obvious $k_z$
influence on its in-plane dispersion is observed as well. Therefore,
unlike the strong $k_z$ dependence of $\alpha$ and $\beta$, the weak
$k_z$ dependence of $\gamma$ is consistent with the nature of
$d_{x^2-y^2}$ and $d_{xy}$, which are more in-plane than the other
$3d$ orbitals. Note that, $\gamma$ is more intensive in both
polarization geometries near Z, likely due to matrix element effects
related to its distribution along the crystallographic $c$ axis.

The band structure and orbital characters around Z are shown in
Fig.~\ref{z}. The $\alpha$ and $\beta$ bands cross $E_F$ along both
the Z-A and Z-R directions, and thus form two hole pockets around
the Z point. The polarization dependence of the spectra near Z are
similar to that near $\Gamma$. Therefore, around the zone center,
all the bands  keep the same spatial symmetry at different $k_z$'s.

\subsection{Orbital characters of bands around the zone corner}

\begin{figure}
\centerline{\includegraphics[width=8.7cm]{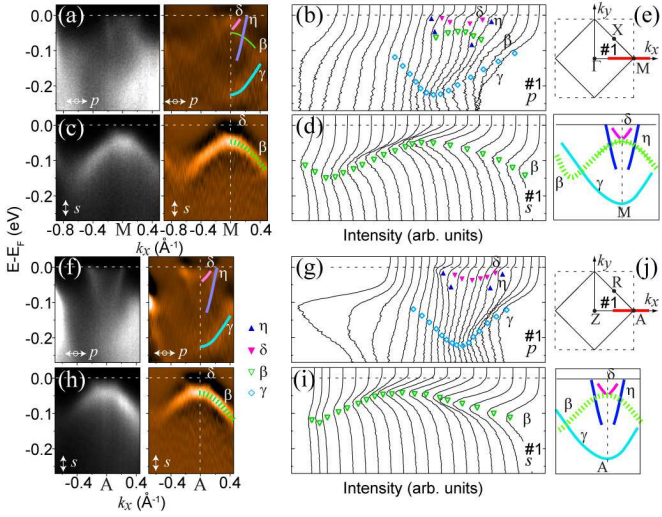}} \caption{(Color
online) The polarization dependent APRES data around the zone
corner. (a) $I(k,\omega)$ and $\partial^2
I(k,\omega)/\partial\omega^2$ map taken with the $p$ geometry and
51~eV photons across M along the red thick line in the inset of
panel e. (b) The EDCs for the data in panel a. (c) and (d) are the
same as in panels a and b, but taken under the $s$ geometry. (e) The
sketch of the band structure in panels a and c. (f)-(j) are the same
as in (a)-(e), but taken with 38~eV photons across A along the red
thick line in the inset of panel j. } \label{m}
\end{figure}

The polarization dependent photoemission data around the zone corner are
plotted in Fig.~\ref{m}. The electron-like $\delta$ band is degenerate with
$\beta$ at about 40~meV below $E_F$ at M.   The $\delta$ band is much more
pronounced in the $p$ geometry, while $\beta$ is very strong in the $s$
geometry. On the other hand, some residual spectral weight of $\beta$ could
still be observed in the $p$ geometry at M, and some infinitesimal $\delta$
band spectral weight could be observed in the $s$ geometry at both M and A,
indicating a light mixing between $\beta$ and $\delta$. If we assume that the
orbital character of $\beta$ does not change abruptly, its main composition can
be then attributed to the odd $d_{yz}$ orbital, based on its $d_{yz}$ orbital
character around $\Gamma$. In Fig.~\ref{setup}(e), the $\delta$ band is
attributed to the even $d_{xz}$ orbital, which is consistent with its mainly
even character observed here.

The $\eta$ band disperses over 150~meV below $E_F$, which is very close to the
$\gamma$ band at about -220~meV. This pair of bands ($\gamma$ and $\eta$) was
previously predicted to be of $d_{xy}$ with odd symmetry [Fig.~\ref{setup}(e)].
However surprisingly, both of them are actually even and only appear in the $p$
geometry around the zone corner [Figs.~\ref{m}(a) and ~\ref{m}(f)].

\begin{figure}
\centerline{\includegraphics[width=8.7cm]{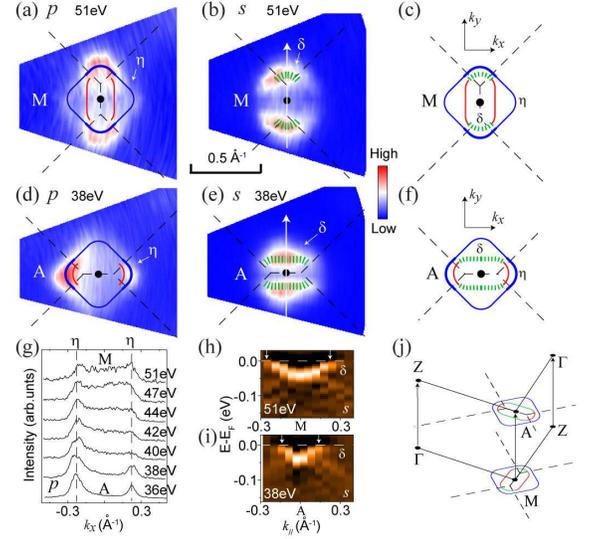}} \caption{(Color
online) Fermi surface around the zone corner. (a) and (b) are the
photoemission intensity maps taken with 51~eV photons around M  in
the $p$ and $s$ geometry respectively. (c) The sketch of the Fermi
surface sheets observed in panels a and b. (d)-(f) are the same as
in (a)-(c), but taken with 38~eV photons around  A. (g) The photon
energy dependence of the MDCs at $E_F$ in the $p$ geometry along the
$\Gamma$(Z)-M(A) direction. (h) and (i) $\partial^2
I(k,\omega)/\partial\omega^2$ map obtained by merging selected EDCs
in the $s$ geometry along the white arrows in panels b and e,
respectively. (j) The illustration of the Fermi surface at A and M
in the 3D Brillouin zone. }\label{mmap}
\end{figure}

To further settle the orbital characters around the zone corner,
Figs.~\ref{mmap}(a)-\ref{mmap}(f) shows the photoemission intensity map around
the M and A point by rotating the sample azimuthally. Obviously, multiple
disconnected parts of the Fermi surface are either enhanced or suppressed by
the $p$ or $s$ geometries, which again demonstrates the multi-orbital nature of
the Fermi surface. Although the polarization selection rules of different
orbitals in Table~\ref{t1} are not strict when the cuts are not exactly along
the high symmetry direction, they still manifest themselves effectively, for
example, as strong intensity modulations along Fermi surface sheets composed of
sections with opposite spatial symmetries. Combining the Fermi surface parts
observed in both the $p$ and $s$ geometries, one obtains two electron pockets
around the M and A points [Figs.~\ref{mmap}(c) and ~\ref{mmap}(f)]. The inner
electron pocket is contributed by the $\delta$ band. If we assume $\delta$ to
be the $\delta_{xz}$ band along $\Gamma$-M in Fig.~\ref{setup}(e), its
horizontal ($k_x$) sections are made of the $d_{xz}$ orbital, which could be
thus observed in the $p$ geometry [the red solid line in Figs.~\ref{mmap}(a)
and \ref{mmap}(d)]. Since the four-fold and translational symmetries of the
crystal would result in the rotating distribution of the $d_{xz}$ and $d_{yz}$
orbitals, the vertical ($k_y$) sections of $\delta$ are made of the $d_{yz}$
orbital, which could be thus observed in the $s$ geometry [the green dashed
line in Figs.~\ref{mmap}(b) and (e)]. Moreover, the elliptical shape of the
$\delta$ pocket rotates 90$^{\circ}$ from M to A, which is consistent with the
3D character of the Brillouin zone [Fig.~\ref{mmap}(j)]. As shown in
Figs.~\ref{mmap}(h) and \ref{mmap}(i), the distance between the two Fermi
crossings of the $\delta$ band along the vertical ($k_y$) direction shrinks
significantly from M to A. Therefore, the alternating even and odd orbital
nature and strong $k_z$ dependence of $\delta$ electron pocket further supports
its $d_{xz}$ and $d_{yz}$ orbital characters.

The outer square-like pocket, as shown by the blue solid line in
Figs.~\ref{mmap}(a) and \ref{mmap}(d), is from the $\eta$ band. It is even and
thus could be only observed in the $p$ geometry. Among the three even orbitals,
we could first exclude the $d_{xz}$ orbital in the $\eta$ pocket, otherwise,
one would have observed the corresponding $d_{yz}$ component of $\eta$ along
$k_y$ in the $s$ geometry, due to the four-fold symmetry of the sample.
Furthermore, the $\eta$ Fermi pocket is particularly intense in the Z-A or Z-M
direction, as indicated by the thicker blue lines in Figs.~\ref{mmap}(a) and
\ref{mmap}(d). The rotation of these intense sections from A to M suggests the
existence of orbital mixing rather than simple matrix element effect.
Therefore, the $\eta$ band should be the mixture of $d_{x^2-y^2}$ and $d_{z^2}$
orbitals. As shown in Fig.~\ref{mmap}(g), the peak intensity of $\eta$ show
asymmetry with respect to the A point. Similarly, in Fig.~\ref{mmap}(d), the
intense section of $\eta$ in the first Brillouin zone is more intensive than
that in the extended Brillouin zone on the right side. This contradicts the
behavior of the $d_{x^2-y^2}$ orbital, as the matrix element or photoemission
intensity of $d_{x^2-y^2}$ should became stronger as the in-plane momentum
became larger. We thus conclude that the intense sections of $\eta$ are consist
of the $d_{z^2}$ orbital over these momentum regions. On the other hand,
considering the facts that the $\eta$ electron pocket shows no $k_z$ dependence
as the corresponding MDCs peak positions do not change with the photon energies
in Fig.~\ref{mmap}(g), the $\eta$ pocket is mostly made of the more in-plane
$d_{x^2-y^2}$ orbital.

The orbital assignment of the $\eta$ pocket could naturally explain the orbital
character of the $\gamma$ band near the zone corner. Since $\gamma$ is actually
the same band as $\eta$ in the unfolded Brillouin zone, they are thus
degenerate and have the same orbital characters at M \cite{Kuroki,Graser1}.
Therefore, the $\gamma$ band is also dominated by $d_{x^2-y^2}$ at the zone
corner (possibly some $d_{z^2}$), which indicates that the $d_{x^2-y^2}$
component of $\gamma$ grows from $\Gamma$ to M.

\section{Discussion}

\begin{figure}
\centerline{\includegraphics[width=8.7cm]{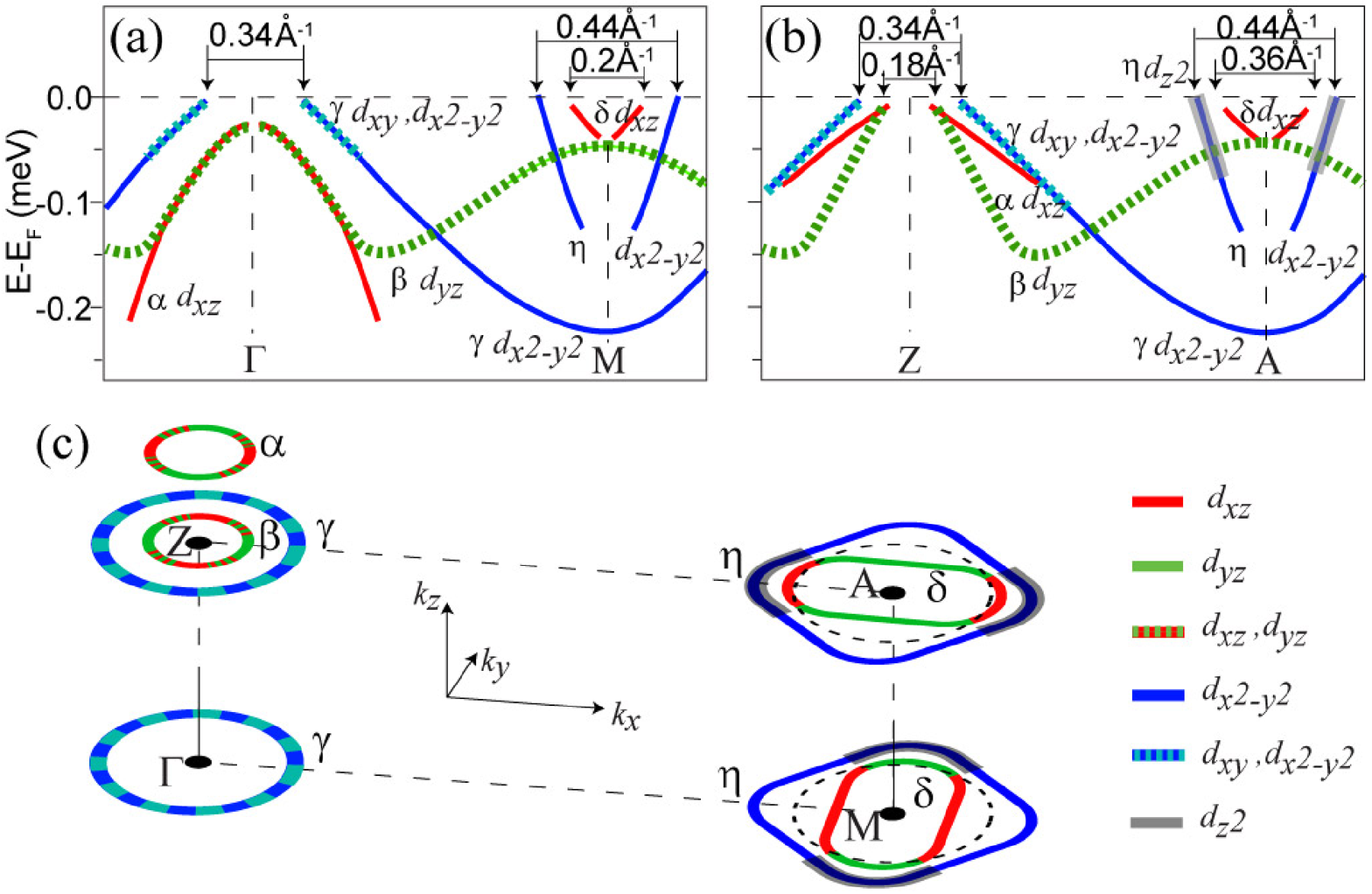}} \caption{(Color
online) The summary of the orbital characters of low energy
electronic structure. (a) and (b) The orbital characters of low
energy electronic structure along $\Gamma$-M, and Z-A respectively.
The solid and dashed lines represent the bands observed in the $p$
and the $s$ geometry respectively. (c) The illustration of the
orbital characters on the Fermi surface sheets at $\Gamma$, M, Z,
and A. The almost overlapping $\alpha$ and $\beta$ Fermi surface
sheets at Z are separated for a better illustration. The black
dashed circles around M and A are the duplicates of the $\gamma$
Fermi surface sheets shifted from $\Gamma$ and Z. The slight mixing
between $\delta$ and $\beta$ near the zone corner is not shown
though.} \label{sum}
\end{figure}

Based on the above analysis, the orbital characters of the low-energy
electronic structure of BaFe$_{1.85}$Co$_{0.15}$As$_2$ are summarized in
Fig.~\ref{sum}. The bands with $d_{xz}$ and $d_{yz}$ orbitals, including
$\alpha$, $\beta$ and $\delta$, are well consistent with the band calculation
in Fig.~\ref{setup}(e). The in-plane dispersion of the $\alpha$ band varies
much more strongly with $k_z$ than that of the $\beta$ band. This difference in
the $k_z$ dependencies between $\alpha$ and $\beta$ is consistent with the
density function theory calculations in BaFe$_2$As$_2$ \cite{LDA2} and previous
ARPES studies \cite{YZhangBK, BaCo3D}. In particular, in the 3D  model of
BaFe$_2$As$_2$ \cite{Graser2}, the sizes of the hole pockets at Z are larger
than that at $\Gamma$, which quantitatively agrees with our results as well.
However, our experimental result seems  not to support the orbital character
change from  Z-A to Z-R predicted in Ref.\cite{Graser2}, which states that the
$\alpha$ and $\beta$ would be correspondingly composed of $d_{xy}$ and
$d_{x^2-y^2}$ orbital along the Z-R direction.

The most remarkable findings here are the orbital characters of the $\gamma$
and $\eta$ bands. They are dominated by the $d_{x^2-y^2}$ orbital around the
zone corner, and  $\gamma$ is a mixture of $d_{xy}$ and $d_{x^2-y^2}$ at the
zone center. However, in most band calculations, the  $\gamma$ and $\eta$ bands
are proposed to be purely  $d_{xy}$, and the $d_{x^2-y^2}$ states are far from
$E_F$. Our results disprove this picture. Furthermore, the band top of the
$\gamma$ band is found to be higher than those of the $\alpha$ and $\beta$ at
the zone center, which is also inconsistent with the band calculations shown in
Fig.~\ref{setup}(e). Considering the 2D character of the the $\gamma$ and
$\eta$ bands, the $d_{xy}$ and $d_{x^2-y^2}$ orbitals should interact less with
states outside the Fe layer (such as As $4p_z$ states) than the other $3d$
orbitals. Therefore, our results indicate that  the energy of the $d_{xy}$ and
$d_{x^2-y^2}$ orbitals may not be calculated correctly, and electron-electron
correlations may be strongly  orbital dependent and result in the deviation
from a simple local density approximation (LDA) calculation.

The pairing in the superconducting state can be strongly tied to the
simultaneous existence of both Fermi surface sheets between the zone center and
zone corner in iron-based superconductors. It has been suggested that  the
pairing is the strongest when the nesting condition between these two Fermi
surface sheets are reached\cite{HDing, BaCo}. However, in Fig.~\ref{sum},
$\alpha$ and $\beta$ show Fermi crossings with the distance about
0.18$\AA^{-1}$ at the Z point , which is still smaller than the 0.2$\AA^{-1}$
of the $\delta$ band at the M point. Therefore, the $\alpha$ and $\beta$ bands
do not nest well with the $\delta$ band. The $\gamma$ and $\eta$ bands form one
hole pocket and one electron pocket respectively with a cylinder-like shape in
3D Brillouin zone due to their 2D character and they are clearly not nested to
each other as demonstrated by the black dashed circles in Fig.~\ref{sum}(c). If
we include the $k_z$ dependence, the size of the $\delta$ electron pocket
changes from 0.2$\AA^{-1}$ to 0.36$\AA^{-1}$, passing the 0.34$\AA^{-1}$ of the
$\gamma$ bands. Therefore, the only possible nesting is between $\gamma$  and
$\delta$ within a small $k_z$ momentum window. It is hard to argue that this
type of nesting can effectively enhance pairing since it takes place at a small
region of $k_z$. In summary, our result does not support that the nesting
condition (at least the strict nesting condition) plays a strong role in
forming Cooper pairs.


\section{Summary}

To summarize, we have carried out a systematic investigation on orbital
characters of BaCo$_{0.15}$Fe$_{1.85}$As$_2$ and analyzed all the possible
orbital characters of the multi-bands based on their strong
polarization-dependent photoemission response. As recapitulated in
Fig.~\ref{sum}, we found that although the main band structure is qualitatively
consistent with the prediction of the present 2D band model, there are
important differences between the current theories and our experimental
results. The distribution of the $d_{xz}$, $d_{yz}$, and $d_{z^2}$ orbital
character  agree well with the theories, but the bands predicted to be $d_{xy}$
orbital show a strong mixture with $d_{x^2-y^2}$ orbital around the zone center
and almost pure $d_{x^2-y^2}$ orbital around the zone corner. The discrepancy
between the calculated band structures and our experimental results suggests
that orbital dependent electron-electron correlations play an important role.
We also study the 3D character of the electronic structure  and show the strong
$k_z$ dependence of the Fermi surfaces  and in-plane dispersion of certain
bands, which proves that the 3D character is also strongly orbital dependent.
Our results  lay out a comprehensive picture of the orbital identities of the
multi-band electronic structures, and provide explicit ingredients  for
constructing the theory of iron-based superconductors.

\section{Acknowledgments}

This work was supported by the NSFC, MOE, MOST (National Basic Research
Program), STCSM of China, the NSF of US under grant No. PHY-0603759, and HSRC
under proposal 08-A-28.

\end{document}